\documentclass[journal]{IEEEtran}
\usepackage[normalem]{ulem}
\usepackage[noadjust]{cite}
\usepackage{bbm}
\usepackage{array}
\usepackage{dcolumn}
\usepackage{epsfig}
\usepackage[intlimits]{amsmath}
\usepackage{amsmath, amsfonts, yhmath, bm}
\usepackage{amssymb}
\usepackage{psfrag}
\usepackage{color,soul}
\usepackage[dvipsnames]{xcolor}
\usepackage[normalem]{ulem}
\usepackage{enumerate}
\usepackage{stackengine}
\usepackage[noadjust]{cite}
\usepackage{graphicx}
\usepackage{caption}
\usepackage[font=footnotesize]{subcaption}
\usepackage{multirow}
\usepackage{cite}
\usepackage[font=scriptsize]{caption}
\usepackage{etoolbox}
\usepackage{tcolorbox}
\usepackage{float}
\usepackage{enumitem}
\usepackage{dsfont}
\newcommand {\myvec}[1] {{\mbox{\boldmath $#1$}}}
\newcommand {\mymat}[1]  {{\mbox{\boldmath $#1$}}}

\DeclareMathAlphabet      {\mathbfit}{OML}{cmm}{b}{it}

\newcommand{\etal}{\textit{et al.}}

\newcommand {\mS} {\mymat{S}}
\newcommand {\A} {\mymat{A}}
\newcommand {\mUpsilon} {\mymat{\Upsilon}}
\newcommand {\G} {\mymat{G}}

\newcommand {\hG} {\widehat{\G}}

\newcommand {\Omeg} {\mymat{\Omega}}
\newcommand {\U} {\mymat{U}}

\newcommand {\B} {\mymat{B}}

\newcommand {\hB} {\widehat{\B}}
\newcommand {\hS} {\widehat{\mS}}

\newcommand {\D} {\mymat{D}}

\newcommand {\mGamma} {\mymat{\Gamma}}

\newcommand {\Ep} {\mymat{\mathcal{E}}}

\renewcommand {\O} {\textrm{O}}
\renewcommand {\P} {\mymat{P}}

\newcommand {\R} {\mymat{R}}

\newcommand {\I} {\mymat{I}}
\newcommand {\X} {\mymat{X}}
\newcommand {\Y} {\mymat{Y}}
\newcommand {\V} {\mymat{V}}

\newcommand {\ue} {\myvec{e}}

\newcommand {\ux} {\myvec{x}}

\newcommand {\hus} {\widehat{\us}}
\newcommand {\hs} {\widehat{s}}

\newcommand {\uv} {\myvec{v}}
\newcommand {\ut} {\myvec{t}}

\newcommand {\us} {\myvec{s}}

\newcommand {\uy} {\myvec{y}}

\newcommand {\Rset} {\mathbb{R}}

\newcommand {\Eset} {\mathbb{E}}

\newcommand {\Tr} {\text{\normalfont Tr}}
\newcommand {\tps} {\rm{T}}
\newcommand {\dCov} {\mathcal{V}}
\newcommand {\dCor} {\mathcal{R}}

\DeclareMathOperator*{\argmin}{argmin}

\DeclareMathOperator{\rank}{rank}
\DeclareMathOperator{\SDC}{SDC}

\newcommand\independent{\protect\mathpalette{\protect\independenT}{\perp}}
\def\independenT#1#2{\mathrel{\rlap{$#1#2$}\mkern2mu{#1#2}}}

\def\comment#1{}

\newcommand{\stkout}[1]{
	\color{red}\ifmmode\text{\sout{\ensuremath{#1}}}\else\sout{#1}\fi\color{black}}

\newcommand{\addra}{}
\newcommand{\delra}{\comment}

\begin{document}

\title{Blind Determination of the Number of Sources Using Distance Correlation}

\author{Amir Weiss and Arie Yeredor
	
	\thanks{The authors are with the School of Electrical Engineering, Faculty of Engineering, Tel-Aviv University,
		P.~O.~Box 39040, Tel-Aviv 69978, Israel, e-mail:
		amirweiss15@gmail.com, arie@eng.tau.ac.il }
	\thanks{The first author would like to thank the Yitzhak and Chaya Weinstein Research Institute for Signal Processing for a fellowship.}
}

\maketitle

\begin{abstract}
A novel blind estimate of the number of sources from noisy, linear mixtures is proposed. Based on Sz{\'e}kely \etal's \textit{distance correlation} measure, we define the Sources' Dependency Criterion (SDC), from which our estimate arises. Unlike most previously proposed estimates, the SDC estimate exploits the full independence of the sources and noise, as well as the non-Gaussianity of the sources \delra{vs.\ }\addra{(as opposed to} the Gaussianity of the noise\addra{)}, via implicit use of high-order statistics. This leads to a more robust, resilient and stable estimate w.r.t.\ the mixing matrix and the noise covariance structure. Empirical simulation results demonstrate these virtues, on top of superior performance in comparison with current state of the art estimates.
\end{abstract}

\vspace{-0.35cm}
\section{Introduction}\label{sec:intro}
The fundamental problem of determining the number of sources from noisy measurements of their linear mixtures has been ubiquitous in signal processing research for decades. This is mainly because correct determination of the model order is a necessary preliminary step in many classical problems in signal processing, such as direction-of-arrival estimation (e.g., \cite{zhou2018off,bakhshi2018role,jiang2017cramer}), blind source separation using Independent Component Analysis (ICA, e.g., \cite{zarzoso2010robust,li2011application,ablin2018faster}) and signal decoding in multiple-input multiple-output wireless systems (see \cite{gesbert2003theory} and references therein), to name but a few.

Many solutions to this problem from various approaches have been proposed in the literature so far, such as the well-known Akaike Information Criterion (AIC) and Minimum Description Length (MDL) \cite{wax1985detection}, Random Matrix Theory (RMT)-based \cite{kritchman2009non}, Second ORder sTatistic of the Eigenvalues
(SORTE) \cite{he2009efficient,he2010detecting}, the recently proposed Bayesian information criterion variant \cite{huang2016bayesian}, mean squared eigenvalue error \cite{beheshti2018number}, and many others \cite{suzuki2000detection,wu2002determination,chung2004detection,choqueuse2008blind,tu2010study,rezaie2015determination,bazzi2016detection}. However, all these solutions are heavily based on an assumption of spatial-whiteness of the additive noise, which essentially leads to a (matrix) rank estimation problem. Thus, to the best of our knowledge, previously proposed methods eventually make explicit use of the eigenvalues of the measurements' empirical correlation matrix for the final estimation rule.

In this work, we address the problem of \textit{blind} determination of the number of sources, where only few, basic assumptions are made, leaving the model general and suitable for a wider range of applications. In contrary to previously proposed methods, our estimate is \textit{not} directly based on the empirical correlation matrix' eigenvalues, and implicitly incorporates high-order statistics, relying on the Gaussianity of the noise vs.\ the non-Gaussianity of the sources. As a result, our estimate is indifferent to the spatial correlation of the noise, and is considerably more robust and resilient in comparison with other dominant, state-of-the-art estimates. Since our proposed solution is based on the (still) less known distance covariance measure, the following section is devoted to a presentation of its definition and some of its interesting, important properties.
\vspace{-0.35cm}
\section{Distance Covariance and Distance Correlation}\label{sec:distancecovariance}
\textit{Distance covariance} (dCov), introduced by Sz{\'e}kely \etal\  \cite{szekely2007measuring}, is a measure which quantifies the dependence between two random vectors, not necessarily of the same dimension.

More formally, let $\ux\in\Rset^{M\times 1}$ and $\uy\in\Rset^{N\times 1}$ be two random vectors with finite first moments\delra{ (from here on throughout this section)}. The dCov between $\ux$ and $\uy$ is the nonnegative number $\dCov(\ux,\uy)$ defined by
\begin{align}\label{dcovdef}
	\dCov^2(\ux,\uy)&\triangleq\|\varphi_{\scriptsize{\ux,\uy}}(\ut,\us)-\varphi_{\scriptsize{\ux}}(\ut)\varphi_{\scriptsize{\uy}}(\us)\|^2\nonumber\\
	&\triangleq\int_{\Rset^{M+N}}{\frac{|\varphi_{\scriptsize{\ux,\uy}}(\ut,\us)-\varphi_{\scriptsize{\ux}}(\ut)\varphi_{\scriptsize{\uy}}(\us)|^2}{c_Mc_N\|\ut\|_2^{M+1}\|\us\|_2^{N+1}}\text{d}\ut\text{d}\us},
\end{align}
where $c_d\triangleq\frac{\pi^{(1+d)/2}}{\Gamma\left((1+d)/2\right)}$, $\Gamma(\cdot)$ is the complete Gamma function (e.g., \cite{andrews1992special}), $\varphi_{\scriptsize{\ux}}(\ut),\varphi_{\scriptsize{\uy}}(\us)$ and $\varphi_{\scriptsize{\ux,\uy}}(\ut,\us)$ denote the characteristic functions of $\ux,\uy$ and $\left(\ux,\uy\right)$, resp., and $\|\cdot\|_2$ denotes the $\ell^2$ norm. Similarly, \textit{distance variance} (dVar) is defined as the square root of
\begin{equation}\label{dVardef}
\dCov^2(\ux)\triangleq\dCov^2(\ux,\ux)\triangleq\|\varphi_{\scriptsize{\ux,\ux}}(\ut,\us)-\varphi_{\scriptsize{\ux}}(\ut)\varphi_{\scriptsize{\ux}}(\us)\|^2.
\end{equation}
The \textit{distance correlation} (dCor) between $\ux$ and $\uy$ is the nonnegative coefficient $\dCor(\ux,\uy)$ defined by
\begin{equation}\label{dCordef}
	\dCor^2(\ux,\uy)\triangleq\begin{cases}
	\frac{\dCov^2(\small{\ux,\uy})}{\sqrt{\dCov^2(\small{\ux)}\dCov^2(\small{\uy})}}, & \dCov^2(\ux)\dCov^2(\uy)>0,\\ 
	0, & \dCov^2(\ux)\dCov^2(\uy)=0.
	\end{cases}
\end{equation}
An important property of dCor is the following (e.g., \cite{szekely2007measuring}):
\begin{enumerate}[label=P{{\arabic*}}:]
	\item $0\leq\dCor(\ux,\uy)\leq1$, and $\dCor(\ux,\uy)=0$ \textit{if and only if} $\ux\independent\uy$ (which denotes $\ux$ and $\uy$ are statistically independent).
\end{enumerate}
We stress that unlike the classical Pearson's correlation coefficient (e.g., \cite{benesty2009pearson}), which may equal zero even if its arguments are statistically dependent, zero dCor necessarily implies statistical independence of its arguments.

Remarkably, very simple empirical estimates of the distance covariance exist, which do not require direct integration: For an observed random independent, identically distributed (i.i.d.) sample $\left\{(\ux_t,\uy_t)\right\}_{t=1}^T=\left\{\X\in\Rset^{M\times T},\Y\in\Rset^{N\times T}\right\}$ from the joint distribution of $\ux$ and $\uy$, define\vspace{-0.125cm}
\begin{multline}\label{empiricalstats}
\left(\mUpsilon^{(x)}\right)_{t,\tau}\triangleq\|\ux_{t}-\ux_{\tau}\|_2, \; \left(\mUpsilon^{(y)}\right)_{t,\tau}\triangleq\|\uy_{t}-\uy_{\tau}\|_2, \\
 \forall t,\tau\in\{1,\ldots,T\}.
\end{multline}
The \textit{empirical dCov}, then, is the nonnegative number $\dCov_T(\X,\Y)$, defined by
\begin{equation}\label{empiricaldCov}
\dCov^2_T(\X,\Y)\triangleq\frac{1}{T^2}\Tr\left(\P\mUpsilon^{(x)}\P\mUpsilon^{(y)}\right),
\end{equation}
where $\P\triangleq\I_T-\tfrac{1}{T}\bf{1}\bf{1}^{\tps}$ is a projection matrix  ($\I_T$ denoting the $T\times T$ identity matrix and $\bf{1}$ denoting a $T\times 1$ all-ones vector). Accordingly, the \textit{empirical dVar} $\dCov_T(\X)\in\Rset^{+}$ is defined by
\begin{equation}\label{empiricaldVar}
\dCov^2_T(\X)\triangleq\dCov^2_T(\X,\X)=\frac{1}{T^2}\Tr\left(\P\mUpsilon^{(x)}\P\mUpsilon^{(x)}\right),
\end{equation}
and the \textit{empirical dCor} $\dCor_T(\X,\Y)\in[0,1]$ is defined by
\begin{equation}\label{empiricaldCordef}
\dCor_T^2(\X,\Y)\triangleq\begin{cases}
\frac{\dCov_T^2\big({\small{\X,\Y}}\big)}{\sqrt{\dCov_T^2\big(\small{\X}\big)\dCov_T^2\big(\small{\Y}\big)}}, & \dCov_T^2(\X)\dCov_T^2(\Y)>0,\\ 
0, & \dCov_T^2(\X)\dCov_T^2(\Y)=0.
\end{cases}
\end{equation}
Note that the statistic $\dCor_T(\X,\Y)$ may be computed rather simply (in terms of arithmetic operations), which is important in our context for practical considerations. As shown in \cite{szekely2014partial},
\begin{enumerate}[label=P{{\arabic*}}:]
	\setcounter{enumi}{1}
	\item $\dCov_T(\X,\Y)$ and $\dCor_T(\X,\Y)$ converge almost surely (a.s.) to $\dCov(\X,\Y)$ and $\dCor(\X,\Y)$, resp., as $T\rightarrow\infty$.
\end{enumerate}
Note also that according to \cite{huo2016fast}, an unbiased estimate of dCor may be computed in $\mathcal{O}(T\log{}T)$ operations, which makes it even more practical for applications with large sample sizes.

Having established the foundations for our proposed estimate, we now turn to the problem in hand.
\section{Problem Formulation}\label{sec:problemformulation}
Consider the linear, instantaneous noisy ICA model
\begin{equation}\label{modelvec}
	\ux[t]=\A\us[t]+\uv[t]\in\Rset^{L\times 1}, \forall t\in\{1,\ldots,T\},
\end{equation}
which may be written conveniently in matrix form as $\X=\A\mS+\V\in\Rset^{L\times T}$, where $\mS=\left[\us_1\;\cdots\;\us_M\right]^{\rm{T}}\in\Rset^{M\times T}$ denotes a matrix of $M>1$ source signals of length $T$, $\A\in\Rset^{L\times M}$ is a (deterministic) full rank mixing matrix, $\V=\left[\uv_1\;\cdots\;\uv_L\right]^{\rm{T}}\in\Rset^{L\times T}$ denotes a matrix of $L$ additive noise signals (one for each sensor), where we assume $L>M$, and $\X=\left[\ux_1\;\cdots\;\ux_L\right]^{\rm{T}}\in\Rset^{L\times T}$ is the matrix of the observed mixture signals. As in the standard ICA model, the sources $\us_1,\ldots,\us_M\in\Rset^{T\times1}$ (i.e., the rows of $\mS$) are assumed to be mutually statistically independent random processes, associated with unknown distributions, and the mixing matrix $\A$ is assumed to be unknown. However, unlike the common (not necessarily justified) assumption that the number of sources is known, here $M$ is considered to be (deterministic) \textit{unknown}. For notational convenience \textit{only}, we assume \addra{that }all the signals involved are zero mean. We also assume \addra{that }the sources are non-Gaussian and \addra{that }each source is temporally i.i.d. As a scaling convention we assume, without loss of generality, that the spatial covariance of the sources is $\Eset\left[\us[t]\us[t]^{\tps}\right]=\I_M$ since the sources' scales are non-identifiable in this model. Furthermore, we assume that the noise $\uv_1,\ldots,\uv_L\in\Rset^{T\times1}$ from all the sensors (i.e., the rows of $\V$) are temporally-white Gaussian noise processes, statistically independent from all the sources, with an unknown spatial covariance matrix $\Eset\left[\uv[t]\uv[t]^{\tps}\right]\triangleq\R_v\in\Rset^{L\times L}$, where $\R_v$ can be any Positive-Definite (PD) matrix. This completes the definition of our model and the problem in question may be stated concisely as follows:
\tcbset{colframe=black!90!blue,size=small,width=0.49\textwidth,halign=flush center,arc=2mm,outer arc=1mm}
\begin{tcolorbox}[upperbox=visible,colback=white]
\textbf{Problem:} \textit{Given $\X$, determine the number of sources $M$.}
\end{tcolorbox}
\vspace{-0.2cm}
\section{The Sources' Dependency Criterion Estimate}\label{sec:solution}
Our proposed solution approach is based on the ability to \textit{injectively} determine the empirical statistical independence of estimated sources using the empirical dCor. However, in order to put this powerful tool to work in the context of our problem, we first assume that we have at our disposal an ICA algorithm which can be applied to the $L$ mixture signals using any hypothesized number of sources $N$ (``$N$-hypothesis") with $1<N<L$, and provides consistent separation in the following sense: Let
\begin{align}\label{sourcesestimate}
	\hS\hspace{1sp}^{(N)}\triangleq\hB\hspace{1sp}^{(N)}\X&=\hB\hspace{1sp}^{(N)}\A\mS+\hB\hspace{1sp}^{(N)}\V\\ \nonumber
	&\triangleq\hG\hspace{1sp}^{(N)}\mS+\hB\hspace{1sp}^{(N)}\V\in\Rset^{N\times T}
\end{align}
denote the output of the separation algorithm under the $N$-hypothesis, where $\hB\hspace{1sp}^{(N)}\in\Rset^{N\times L}$ and $\hG\hspace{1sp}^{(N)}=\hB\hspace{1sp}^{(N)}\A\in\Rset^{N\times M}$ denote, resp., the estimated separating matrix and the resulting overall mixing-unmixing matrix, all under the same $N$-hypothesis. By ``consistency" we mean that asymptotically (in both SNR and sample size together) perfect separation is obtained for any $N\geq M$, namely $\hG\hspace{1sp}^{(M)}=\mGamma^{(M)}$, and for $N>M$, $\hG\hspace{1sp}^{(N)}$ has $\mGamma^{(N)}$ as its top $M\times M$ block and all-zeros as its lower $(N-M)\times M$ block, where $\{\mGamma^{(N)}\in\Rset^{M\times M}\}_{N=M}^{L}$ are a set of some scaled permutation matrices (which, in general, may differ from one another). We note that some prominent classical ICA algorithms, such as JADE \cite{cardoso1990eigen} or FastICA \cite{hyvarinen1999fast}, enjoy such a consistency property.

Equipped with a consistent ICA algorithm, and observing that due to properties P1 and P2, a.s.
\begin{multline}
\lim_{T\rightarrow\infty}\dCor_T(\us_{m_1},\us_{m_2})=\dCor\left(s_{m_1}[t],s_{m_2}[\tau]\right)=0, \\ \forall 1\leq m_1\neq m_2\leq M, \; \forall t,\tau\in\{1,\ldots,T\},
\end{multline}
we propose the following estimate for the number of sources:
\begin{equation}\label{SDCestimate}
	\widehat{M}_{\text{\scriptsize{SDC}}}\triangleq\argmin_{N\in\{2,\ldots,L-1\}}{\SDC(N)}
\end{equation}
where the Sources' Dependency Criterion / Sources' empirical Distance Correlation (SDC) is defined (for $1<N<L$) as
\begin{equation}\label{SDCdef}
\SDC(N)\triangleq\max_{n\in\{1,\ldots,N\}}{\dCor_T\left(\hus_{n}^{(N)},\hus_{N+1}^{(N+1)}\right)}.
\end{equation}
Put simply, the SDC measures the maximal empirical dCor between each of the $N$ estimated sources under the $N$-hypothesis and the ``new" additional $(N+1)$-th source under the $(N+1)$-hypothesis (i.e., the $(N+1)$-th row of $\hS\hspace{1sp}^{(N+1)}$).

To formally justify and further explain the rationale of the proposed estimate, we shall present an asymptotic (qualitative) analysis of its operation. We start by \delra{denoting}\addra{defining} a few necessary notations. First, we denote the Singular Value Decompositions (SVDs) $\A\triangleq\U_A\D_A\V_A^{\tps}$ and $\R_v\triangleq\U_v\D_v\U_v^{\tps}$, and we assume that the singular values are sorted in a decreasing order on the diagonals of $\D_A$ and $\D_v$. With this, we have
\begin{equation}\label{mixturescovariance}
	\R_x\triangleq\Eset\left[\ux[t]\ux[t]^{\tps}\right]=\U_A\D^2_A\U_A^{\tps}+\U_v\D_v\U_v^{\tps}\triangleq\U_x\D_x\U_x^{\tps}.
\end{equation}
From Weyl's inequality (e.g., \cite{franklin2012matrix}), we have for all $1\leq\ell\leq L$
\begin{equation}\label{Weylinequality}
\left(\D^2_A\right)_{\ell,\ell}+\left(\D_v\right)_{L,L}\leq \left(\D_x\right)_{\ell,\ell} \leq \left(\D^2_A\right)_{\ell,\ell}+\left(\D_v\right)_{1,1}.
\end{equation}
Since $\R_v$ is PD, $\left(\D_v\right)_{\ell,\ell}>0$ for every $1\leq\ell\leq L$. Therefore,
\begin{equation}\label{neweignotations}
\forall\ell\in\{1,\ldots,L\}\hspace{-0.05cm}:\hspace{-0.05cm}\exists\tilde{\sigma}^2_{\ell}\in\Rset^{+}:\left(\D_x\right)_{\ell,\ell} \triangleq \left(\D^2_A\right)_{\ell,\ell}+\tilde{\sigma}^2_{\ell},
\end{equation}
such that $\left(\D_v\right)_{L,L}\leq\tilde{\sigma}^2_{\ell}\leq\left(\D_v\right)_{1,1}$ for all $1\leq\ell\leq L$. Notice that $\left(\D_x\right)_{\ell,\ell}=\tilde{\sigma}^2_{\ell}$ for $M+1\leq\ell\leq L$, since $\rank(\A)=M$. With these notation, we assume
\begin{enumerate}[label=A{{\arabic*}}:]
	\item $\left(\D_v\right)_{1,1}\hspace{-0.05cm}\ll\hspace{-0.05cm}\left(\D^2_A\right)_{M,M}\Rightarrow1\leq\ell\leq L:\tilde{\sigma}^2_{\ell}\ll\left(\D^2_A\right)_{M,M}$, i.e., high SNR.
	\item The sample size $T$ is (finite but) ``large enough" such that we may approximate $\tfrac{1}{T}\X\X^{\tps}\approx\R_x$, $\dCor_T(\cdot,\cdot)\stackrel{\text{P}2}{\approx}\dCor(\cdot,\cdot)$.
	\item Approximately ``successful" operation of the separation algorithm for $N\geq M$ under A1 and A2:
		\begin{gather*}
			N=M: \hG\hspace{1sp}^{(M)}=\mGamma^{(M)}+\Ep^{(M)}\stackrel{\text{A}1,\text{A}2}{\approx}\mGamma^{(M)},\\ \nonumber
			N>M: \hG\hspace{1sp}^{(N)}=\begin{bmatrix}
			\mGamma^{(N)}\\
			\O
			\end{bmatrix}+\Ep^{(N)}\stackrel{\text{A}1,\text{A}2}{\approx}\begin{bmatrix}
			\mGamma^{(N)}\\
			\O
			\end{bmatrix},\nonumber
		\end{gather*}
		where $\{\Ep^{(N)}\in\Rset^{N\times L}\}$ denote estimation error matrices.
	\item ``Poor" operation of the separation algorithm for $N<M$: When $N<M$ the resulting $\hG\hspace{1sp}^{(N)}$ is generally a ``non-separating" matrix. At least, in particular,
	\begin{gather*}
		N<M:\exists i_1,i_2: \left(\hG\hspace{1sp}^{(N)}\right)_{i_1,N+1},\left(\hG\hspace{1sp}^{(N+1)}\right)_{i_2,N+1}\neq0.
	\end{gather*}
	\item Elements of the estimated $\hB\hspace{1sp}^{(N)}$ are generally non-zeros. In particular, for $N>M$, the matrix $\Omeg^{(N)}\triangleq\hB\hspace{1sp}^{(N)}\R_v\left(\hB\hspace{1sp}^{(N+1)}\right)^{\tps}\in\Rset^{N\times N+1}$ satisfies
	\begin{gather*}
		\exists n\in\{M+1,\ldots,N\}:\left(\Omeg^{(N)}\right)_{n,N+1}\neq0.
	\end{gather*}
\end{enumerate}

We shall now examine the three possible cases of the hypothesis test \eqref{SDCestimate}, which defines our proposed estimate.
\subsection{Case 1: $N$-hypothesis, $1<N<M$}\label{subsec:case1}
Assume the $N$-hypothesis, with $1<N<M$. Therefore, in this case we have
\begin{align}\label{SDCcase1}
\SDC(N)&=\max_{n\in\{1,\ldots,N\}}{\dCor_T\left(\hus_{n}^{(N)},\hus_{N+1}^{(N+1)}\right)}\nonumber\\
&\stackrel{\text{A}2}{\approx} \max_{n\in\{1,\ldots,N\}}{\dCor\left(\hs_{n}^{(N)}[t],\hs_{N+1}^{(N+1)}[t]\right)}\triangleq\varrho_N^2,
\end{align}
where $\varrho_N^2>0$, since $N+1\leq M$, hence the $(N+1)$-th estimated source under the $(N+1)$-hypothesis is (at least partially) linearly ``contained" in one of the $N$ estimated sources under the $N$-hypothesis, by A4.
\subsection{Case 2: $M$-hypothesis}\label{subsec:case2}
Assume the $M$-hypothesis, i.e., the true number of sources. In this case, since the separation algorithm is assumed to be consistent, we have
\begin{align}\label{SDCcase2}
\SDC(M)&=\max_{m\in\{1,\ldots,M\}}{\dCor_T\left(\hus_{m}^{(M)},\hus_{M+1}^{(M+1)}\right)}\nonumber\\
&\stackrel{\text{A}2}{\approx} \max_{m\in\{1,\ldots,M\}}{\dCor\left(\hs_{m}^{(M)}[t],\hs_{M+1}^{(M+1)}[t]\right)}\hspace{-0.05cm}\stackrel{\text{A}1,\text{A}3}{\approx}\hspace{-0.05cm}0,
\end{align}
as $M$ out of the $M+1$ estimated sources under the $(M+1)$-hypothesis must be (noisy versions of) the true sources (due to the consistency of the separation algorithm)\addra{,} and the $(M+1)$-th estimated source is (approximately) a linear combination of noise components \textit{only}, by A3. Thus, asymptotically, we approximately have $\hs_{m}^{(M)}[t]\independent\hs_{M+1}^{(M+1)}[t], \forall m\in\{1,\ldots,M\}$.
\vspace{-0.4cm}
\subsection{Case 3: $N$-hypothesis, $M<N<L$}\label{subsec:case3}
Assume the $N$-hypothesis, with $M<N<L$.
By A3, asymptotically, for every $M<\widetilde{N}\leq L$, the ``spurious" estimated sources $\{\hus_{n}^{(\widetilde{N})}\}_{n=M+1}^{\widetilde{N}}$ are (approximately) linear combinations of noise components only, i.e., $\hus_{n}^{(\widetilde{N})}\stackrel{\text{A}3}{\approx} (\hB\hspace{1sp}^{(\widetilde{N})}\V)^{\tps}\ue_n$ (where $\ue_n$ is the $n$-th column of $\I_{\widetilde{N}}$), and are therefore approximately Gaussian. Using the well-known fact that temporally-white Gaussian signals are non-separable in model \eqref{modelvec} (see, e.g., \cite{yeredor2010blind} and references therein), we assert that A5 is highly likely to hold, hence the estimated spurious source $\hus_{N+1}^{(N+1)}$ would be dependent on at least one estimated spurious source out of $\{\hus_n^{(N)}\}_{n=M+1}^N$ a.s. Therefore, in this case we have
\begin{align}\label{SDCcase3}
\SDC(N)&=\max_{n\in\{1,\ldots,N\}}{\dCor_T\left(\hus_{n}^{(N)},\hus_{N+1}^{(N+1)}\right)}\nonumber\\
&\stackrel{\text{A}2}{\approx} \max_{n\in\{1,\ldots,N\}}{\dCor\left(\hs_{n}^{(N)}[t],\hs_{N+1}^{(N+1)}[t]\right)}\hspace{-0.05cm}\triangleq\hspace{-0.05cm}\varrho_N^2,
\end{align}
where $\varrho_N^2>0$ by virtue of A3 and A5. We note that A4 and A5 are typically quite mild conditions, and may be shown more rigorously to be so. However, the details concerning this claim are out of the scope of this paper and here these conditions are regarded as necessary for a proper operation of the proposed estimate. We also note that although $\{\varrho_N^2\}_{N\ne M}$ were only claimed to be positive, they are typically ``far" from zero, in the sense that $\SDC(M)\ll\varrho_N^2, \forall N\ne M$, as we shall demonstrate empirically in the sequel.

In conclusion of all three cases, asymptotically, we have
\begin{equation}\label{SDCconclusion}
\SDC(N)\approx\begin{cases}
0, & N=M,\\ 
\varrho_N^2, & 1\leq N\neq M\leq L,
\end{cases} \Rightarrow \widehat{M}_{\text{\scriptsize{SDC}}}=M.
\end{equation}
We stress that for any finite SNR and sample size $T$, $\SDC(M)\neq0$ a.s. However, asymptotically $\SDC(M)\rightarrow0$\delra{ and accordingly}\addra{, thus} the \addra{resulting }error probability \delra{of the SDC estimate }approaches zero as well\addra{, implying the consistency of the SDC estimate}. The reasons for this are twofold: The estimate $\hB$ approaches a perfect separating matrix (A3) and the empirical dCor approaches dCor (P2, A2). This assures consistently improving performance as the overall SNR and sample size grow, which is not necessarily true for other, previously proposed estimates in spatially non-white noise scenarios for \textit{any} finite (even if large) SNR.

To summarize, the complete proposed solution algorithm to the problem of estimating the number of sources is as follows:
\tcbset{colframe=black!90!blue,size=small,width=0.49\textwidth,halign=flush center,arc=2mm,outer arc=1mm}
\begin{tcolorbox}[upperbox=visible,colback=white,halign=left]
\textbf{The Proposed Solution Algorithm: SDC Estimation}
\textit{\begin{enumerate}[label=\arabic*.]
	\item Initialization: Obtain $\hS\hspace{1sp}^{(2)}\in\Rset^{2\times T}$;
	\item For every $N_{\text{\normalfont{cand}}}=2,\ldots,L-1$ do:
	\begin{enumerate}
		\item Obtain $\hS\hspace{1sp}^{(N_{\text{\normalfont{cand}}}+1)}\in\Rset^{(N_{\text{\normalfont{cand}}}+1)\times T}$ (e.g., via JADE);
		\item Compute $\SDC(N_{\text{\normalfont{cand}}})$ according to \eqref{SDCdef};
	\end{enumerate}
	\item Determine $\widehat{M}_{\text{\scriptsize{\normalfont{SDC}}}}$ according to \eqref{SDCestimate}.
\end{enumerate}}
\end{tcolorbox}
\section{Simulation Results}\label{sec:simulationresults}\vspace{-0.05cm}
We demonstrate the performance of the proposed SDC estimate according to model \eqref{modelvec} in simulation results of four different scenarios. In the last three, we compare it with the MDL, RMT\footnote{with $\beta=1$ and $\alpha=0.1$} and SORTE estimates\footnote{We do not consider AIC since it is an inconsistent estimate \cite{wax1985detection}.}, which, currently being the leading methods, serve as an appropriate benchmark. All the empirical results are based on $10^3$ independent trials. Unless stated otherwise, the elements of $\A$ were independently drawn at each trial from the standard Gaussian distribution.

First, we consider a scenario of $L=7$ sensors and $M=4$ zero-mean, unit variance Laplace distributed sources with white noise, i.e., $\R_v=\sigma^2\I_L$. Fig.\ 1 presents the SDC cost function value for all the hypotheses, $N$, vs.\ $k$, an index determining the sample size and SNR such that $T=500\cdot k$ and $\sigma^2=-5\cdot k$[dB]. In accordance with our asymptotic analysis, it is seen that the SDC cost function yields a consistent estimate.

Next, we consider a scenario of $L=7$ sensors and $M=3$ zero-mean, unit variance Laplace, Uniform and Rademacher (e.g., \cite{hitczenko1994rademacher}) distributed sources. The noise is ``approximately" white, i.e., $\R_v$ is diagonal with $(\R_v)_{\ell,\ell}=\sigma^2[\text{dB}]+\Delta_{\ell}[\text{dB}]$, where $\sigma^2$ is fixed and $\{\Delta_{\ell}\sim \mathcal{N}(0,\epsilon^2)\}_{\ell=1}^7$ are mutually independent perturbations, with $\epsilon$ symbolizing the deviation from an ``ideal" white-noise model. Fig.\ 2 presents the empirical error probabilities of the estimates vs.\ $\epsilon$ for $\sigma^2=-15[\text{dB}]$ and $T=3000$. Evidently, MDL and RMT are sensitive to deviations from the white-noise model, while SDC and SORTE are more resilient to such deviations. And yet, recall that the SDC is \textit{blind}, so (unlike SORTE) it does not exploit the (valuable) prior assumption of white noise.

In the third scenario we consider the case of non-white uncorrelated noise and one dominant source. In particular, $\R_v$ is diagonal with $(\R_v)_{\ell,\ell}\sim U(\sigma_{0}^2,\sigma_{0}^2+\Delta)\hspace{0.02cm}[\text{dB}]$ (mutually independent). Here, all the sources are equiprobable zero-mean 4-PAM signals, all with unit-variance, except for one with variance $\sigma_s^2$[dB], $L=8$, and the mixing matrix' elements were drawn independently from the standard Uniform distribution. Figs.\ \ref{fig:exp3_for_paper_vs_T} and \ref{fig:exp3_for_paper_vs_sigma} present the average empirical error probability vs.\ $T$, when $\sigma_{0}^2=-15$[dB] is fixed, and vs.\ $1/\sigma_{0}^2$, when $T=2000$ is fixed, resp., where $\Delta=30[\text{dB}], \sigma_s^2=18[\text{dB}]$ and the average is taken over $M\in\{2,\ldots,7\}$\footnote{For SORTE $2\leq M\leq 5$, since it can estimate (only) up to $L-3$ sources.}. Firstly, it is seen that the SDC improves as the SNR and sample size increase. Secondly, asymptotic superiority of the SDC over all the other estimates, which wrongfully assume $\R_v$ is a scaled identity matrix, is evident.\addra{ We stress that in the smaller sample-size regime, the SDC performance may be considerably degraded (as seen in Fig. \ref{fig:exp3_for_paper_vs_T}), possibly due to increased variance in the associated empirical estimates beyond second-order statistics.}

In the last scenario we examine the performance in spatially correlated noise and ``troublesome" mixing conditions. Specifically, $\R_v$ has $\sigma^2$ on its diagonal, $0.1\cdot\sigma^2$ on its sub- and super-diagonals, and zero elsewhere. This structure describes a ``small" spatial correlation between two neighboring sensors (only). Further, after $\A$ was drawn, we substitute (only) $(\D_A)_{M,M}=\sqrt{0.1}$, which is mostly expressed in ``difficult" second-order statistics conditions, and specifically challenges assumption A1, taken in the approximate analysis presented above. Figs.\ \ref{fig:exp4_for_paper_L6M3} and \ref{fig:exp4_for_paper_L10M5} present the empirical error probability vs.\ $1/\sigma^2$ for $L=6, M=3$ and $L=10, M=5$, resp., with zero-mean, unit variance uniformly distributed sources and $T=2000$. Clearly, while other estimates reveal considerable sensitivity to these conditions, the SDC is seen to be stable and exhibits a kind of indifference to ``misleading" mixings and weak noise correlations even in the ``space" domain, when the SNR is sufficiently high.

We note that for all the scenarios presented in this section, similar trends are obtained for different values of $L$ and $M$, and, of course, the accuracy of the SDC estimate (in terms of error probability) is constantly improving with an increasing sample size and SNR, as demonstrated in these scenarios.
\begin{figure}[t!]
	\centering
	\begin{subfigure}[b]{0.22\textwidth}
		\includegraphics[width=\textwidth]{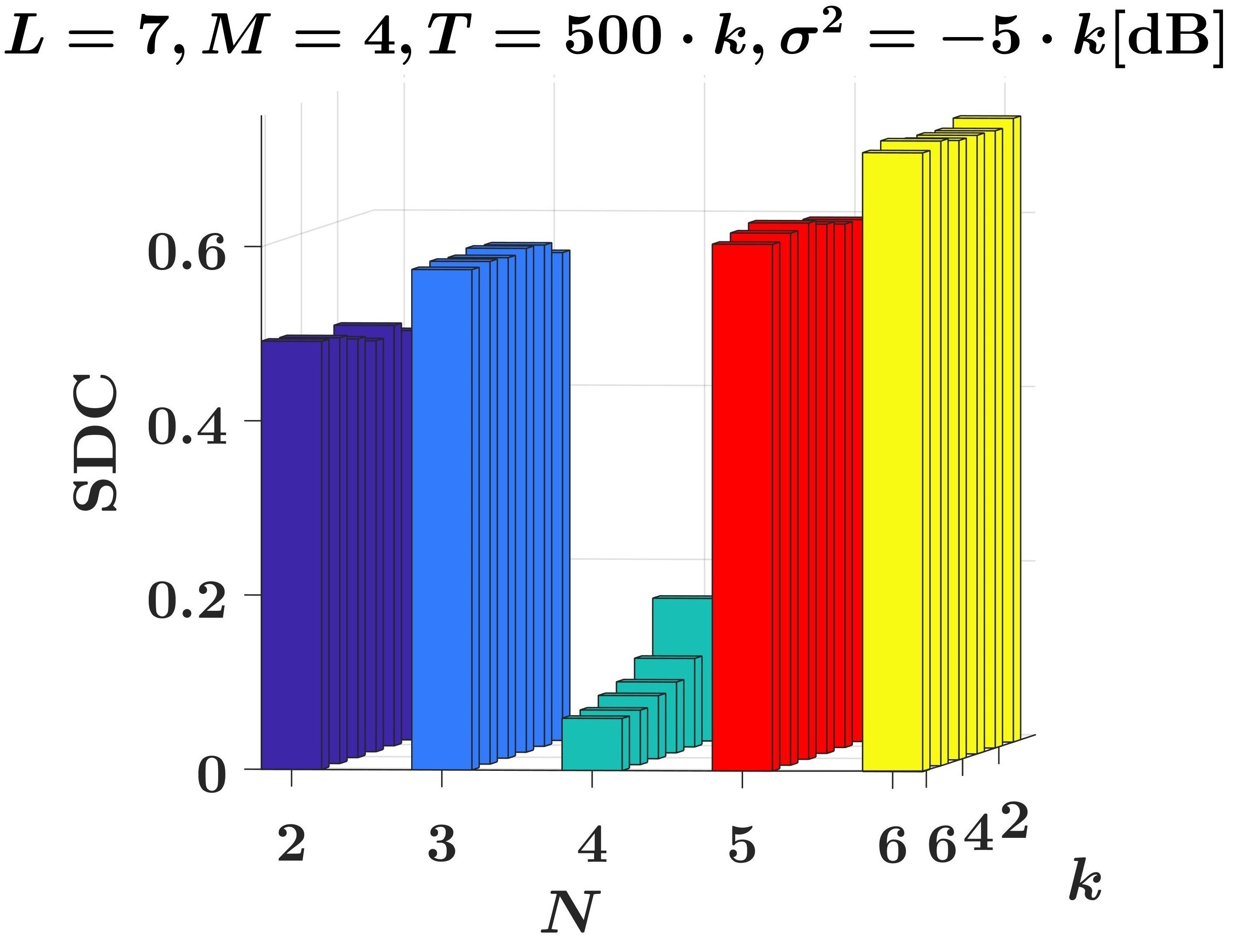}
		\vspace{-0.25cm}
		\label{fig:exp0_for_paper}
	\end{subfigure}%
	~
	\begin{subfigure}[b]{0.22\textwidth}
		\includegraphics[width=\textwidth]{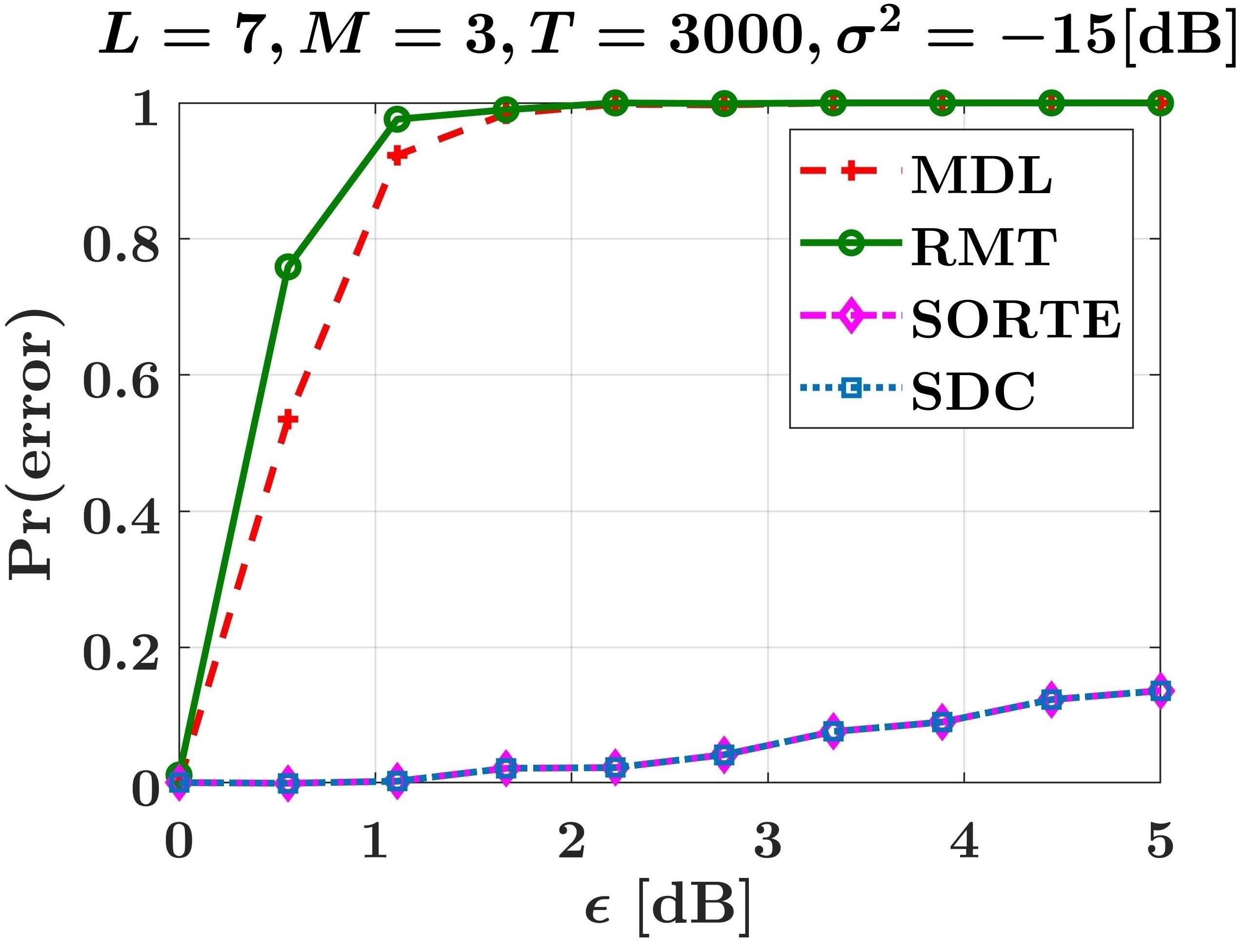}
		\vspace{-0.25cm}
		\label{fig:exp1_for_paper}
	\end{subfigure}\vspace{-0.2cm}
	\quad\quad\quad\quad\quad\quad\quad\quad\quad\quad\quad\quad\quad\quad\quad\quad \scriptsize{Fig.\ 1}\quad\quad\quad\quad\quad\quad\quad\quad\quad\quad\quad\quad\quad\quad \scriptsize{Fig.\ 2} \vspace{-0.1cm}
	\caption{The SDC cost function value vs.\ the $N$-hypothesis vs.\ $k\in\{1,\ldots,6\}$ (index of sample size and SNR) for the first scenario. Indeed, we see that $\SDC(M)\ll\SDC(N)=\varrho_N^2,\forall N\neq M$.\\ \vspace{-0.175cm} \\
		Fig.\ 2: Empirical error probabilities vs.\ $\epsilon$\hspace{0.05cm}[dB] for the second scenario. Note that SORTE chooses from $\{2,3,4\}$, while SDC chooses from $\{2,\ldots,6\}$.}\vspace{-0.15cm}
\end{figure}
\setcounter{figure}{2}
\begin{figure}[t!]
	\vspace{-0.2cm}
	\centering
	\begin{subfigure}[b]{0.22\textwidth}
		\includegraphics[width=\textwidth]{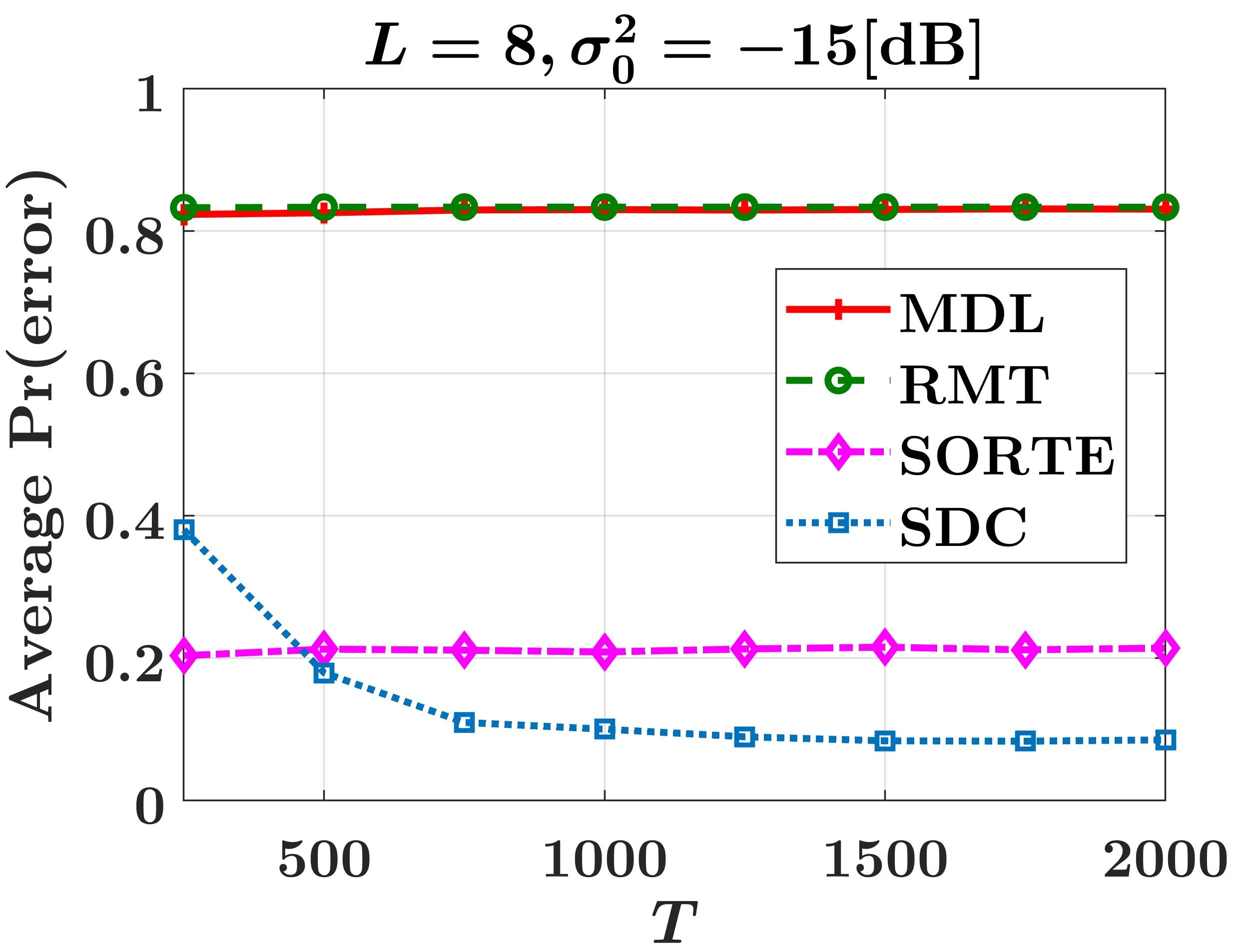}\vspace{-0.15cm}
		\caption{}
		\label{fig:exp3_for_paper_vs_T}\vspace{-0.175cm}
	\end{subfigure}%
	~
	\begin{subfigure}[b]{0.22\textwidth}
		\includegraphics[width=\textwidth]{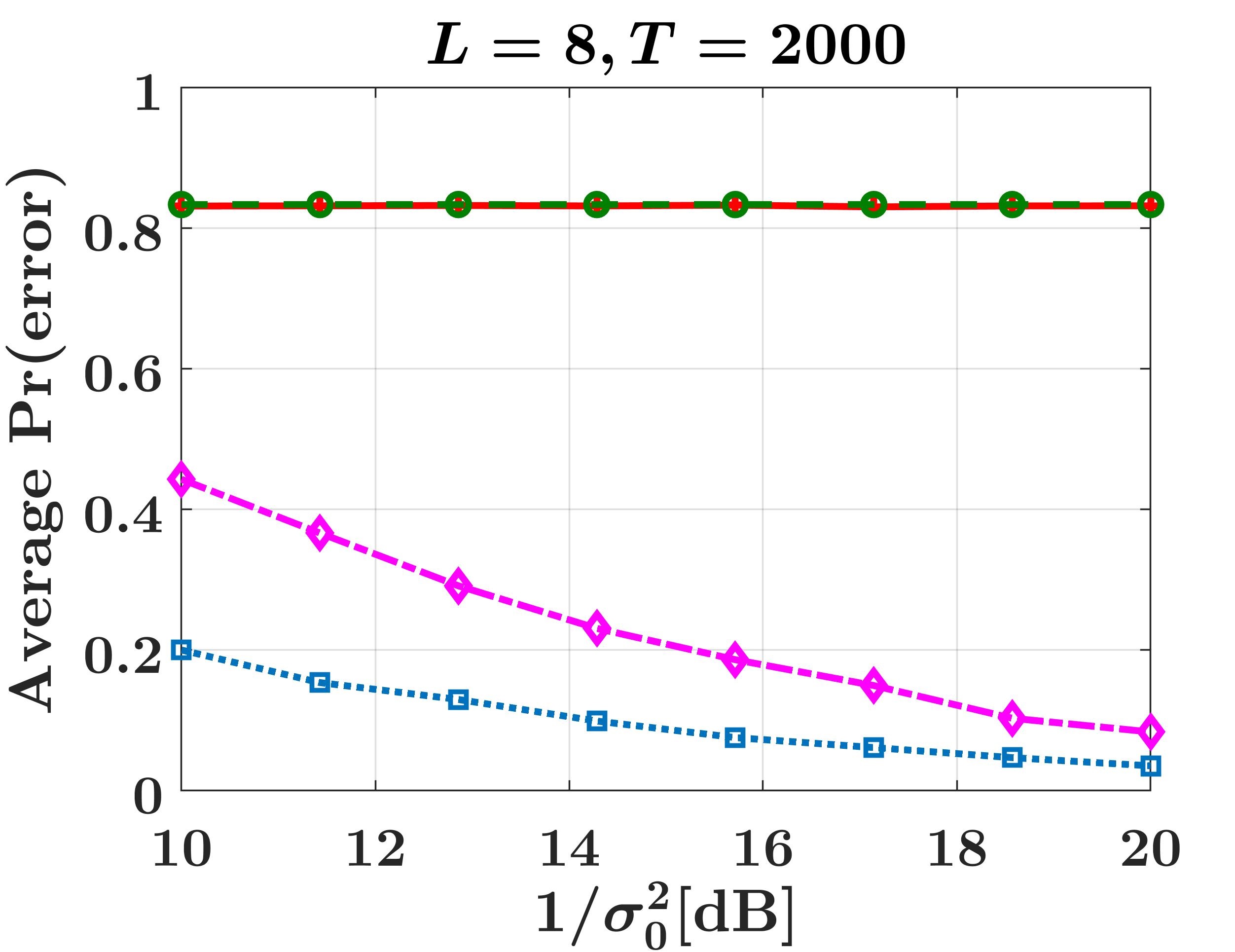}\vspace{-0.15cm}
		\caption{}
		\label{fig:exp3_for_paper_vs_sigma}\vspace{-0.175cm}
	\end{subfigure}
	\caption{Average (over $M$) empirical error probabilities for the third scenario. (a) vs.\ $T$ when $\sigma_{0}^2\hspace{-0.05cm}=\hspace{-0.05cm}-15$[dB] is fixed (b) vs.\ $1/\sigma_{0}^2$ when $T=2000$ is fixed.}\vspace{-0.2cm}
	\label{fig:exp3_for_paper}
\end{figure}
\begin{figure}[t!]
	\vspace{-0.2cm}
	\centering
	\begin{subfigure}[b]{0.22\textwidth}
		\includegraphics[width=\textwidth]{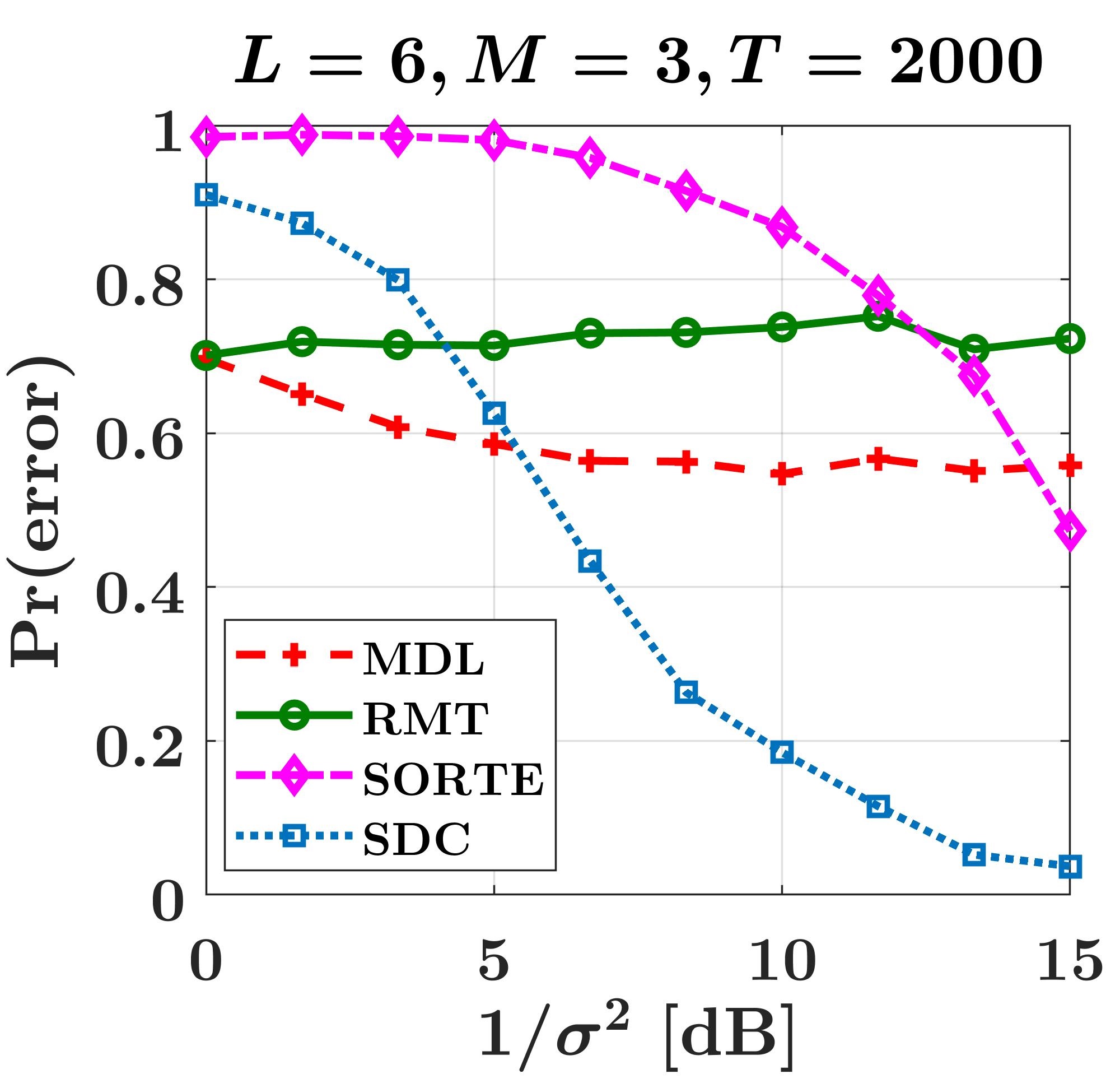}\vspace{-0.15cm}
		\caption{}
		\label{fig:exp4_for_paper_L6M3}\vspace{-0.175cm}
	\end{subfigure}%
	~
	\begin{subfigure}[b]{0.22\textwidth}
		\includegraphics[width=\textwidth]{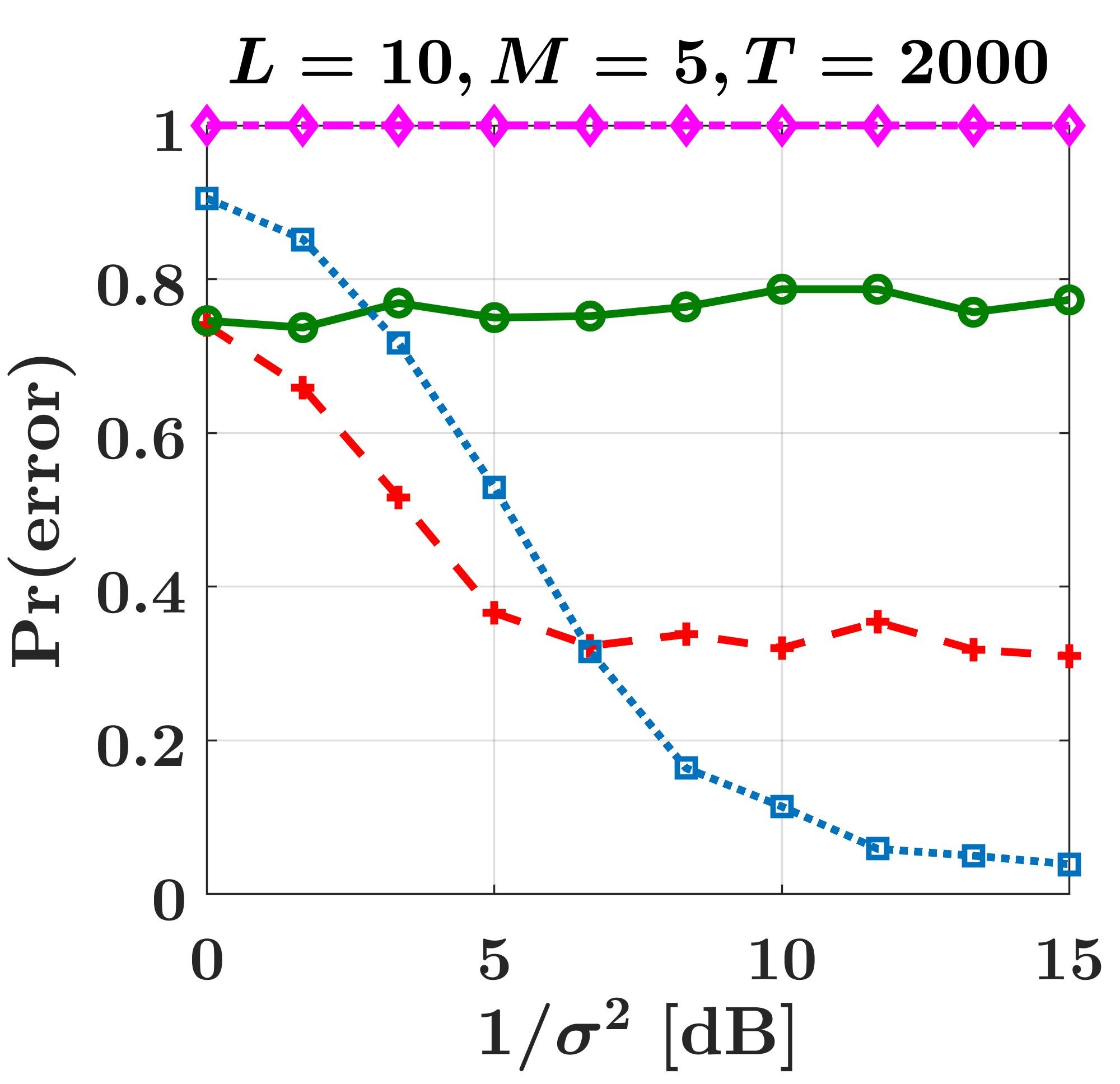}\vspace{-0.15cm}
		\caption{}
		\label{fig:exp4_for_paper_L10M5}\vspace{-0.175cm}
	\end{subfigure}
	\caption{Empirical error probabilities vs.\ $1/\sigma^2$\hspace{0.05cm}[dB] (i.e., SNR) for the fourth scenario with $T=2000$ fixed. (a) $L=6, M=3$ (b) $L=10, M=5$.}\vspace{-0.6cm}
	\label{fig:exp4_for_paper}
\end{figure}
\vspace{-0.375cm}
\section{Conclusion}\label{sec:conclusion}\vspace{-0.075cm}
We presented an algorithm for blind determination of the number of (non-Gaussian) sources from noisy, linear mixtures. The proposed SDC estimate, which arises from the notion of dCor, was shown to be robust and resilient w.r.t.\ the mixing matrix and the noise spatial covariance matrix, which is not assumed to be of any particular structure. Accordingly, it exhibits more stable performance than other estimates when facing deviations from the ideal white-noise model assumption.

\bibliography{Bibfile}

\begin{thebibliography}{10}
\providecommand{\url}[1]{#1}
\csname url@samestyle\endcsname
\providecommand{\newblock}{\relax}
\providecommand{\bibinfo}[2]{#2}
\providecommand{\BIBentrySTDinterwordspacing}{\spaceskip=0pt\relax}
\providecommand{\BIBentryALTinterwordstretchfactor}{4}
\providecommand{\BIBentryALTinterwordspacing}{\spaceskip=\fontdimen2\font plus
\BIBentryALTinterwordstretchfactor\fontdimen3\font minus
  \fontdimen4\font\relax}
\providecommand{\BIBforeignlanguage}[2]{{%
\expandafter\ifx\csname l@#1\endcsname\relax
\typeout{** WARNING: IEEEtran.bst: No hyphenation pattern has been}%
\typeout{** loaded for the language `#1'. Using the pattern for}%
\typeout{** the default language instead.}%
\else
\language=\csname l@#1\endcsname
\fi
#2}}
\providecommand{\BIBdecl}{\relax}
\BIBdecl

\bibitem{zhou2018off}
C.~Zhou, Y.~Gu, Z.~Shi, and Y.~D. Zhang, ``{Off-Grid Direction-of-Arrival
  Estimation Using Coprime Array Interpolation},'' \emph{IEEE Signal Processing
  Letters}, vol.~25, no.~11, pp. 1710--1714, 2018.

\bibitem{bakhshi2018role}
G.~Bakhshi and K.~Shahtalebi, ``{Role of the NLMS Algorithm in Direction of
  Arrival Estimation for Antenna Arrays},'' \emph{IEEE Communications Letters},
  vol.~22, no.~4, pp. 760--763, 2018.

\bibitem{jiang2017cramer}
W.~Jiang and A.~M. Haimovich, ``{Cramer–-Rao Bound for Noncoherent Direction
  of Arrival Estimation in the Presence of Sensor Location Errors},''
  \emph{IEEE Signal Processing Letters}, vol.~24, no.~9, pp. 1303--1307, 2017.

\bibitem{zarzoso2010robust}
V.~Zarzoso and P.~Comon, ``{Robust Independent Component Analysis by Iterative
  Maximization of the Kurtosis Contrast With Algebraic Optimal Step Size},''
  \emph{IEEE Trans. on Neural Networks}, vol.~21, no.~2, pp. 248--261, 2010.

\bibitem{li2011application}
H.~Li, N.~M. Correa, P.~A. Rodriguez, V.~D. Calhoun, and T.~Adali,
  ``{Application of Independent Component Analysis With Adaptive Density Model
  to Complex-Valued fMRI Data},'' \emph{IEEE Trans. on Biomedical Engineering},
  vol.~58, no.~10, pp. 2794--2803, 2011.

\bibitem{ablin2018faster}
P.~Ablin, J.-F. Cardoso, and A.~Gramfort, ``{Faster Independent Component
  Analysis by Preconditioning With Hessian Approximations},'' \emph{IEEE Trans.
  on Signal Processing}, vol.~66, no.~15, pp. 4040--4049, 2018.

\bibitem{gesbert2003theory}
D.~Gesbert, M.~Shafi, D.~shan Shiu, P.~J. Smith, and A.~Naguib, ``{From theory
  to practice: an overview of MIMO space-time coded wireless systems},''
  \emph{IEEE Journal on Selected Areas in Communications}, vol.~21, no.~3, pp.
  281--302, 2003.

\bibitem{wax1985detection}
M.~Wax and T.~Kailath, ``{Detection of signals by information theoretic
  criteria},'' \emph{IEEE Trans. on Acoustics, Speech, and Signal Processing},
  vol.~33, no.~2, pp. 387--392, 1985.

\bibitem{kritchman2009non}
S.~Kritchman and B.~Nadler, ``{Non-Parametric Detection of the Number of
  Signals: Hypothesis Testing and Random Matrix Theory},'' \emph{IEEE Trans. on
  Signal Processing}, vol.~57, no.~10, pp. 3930--3941, 2009.

\bibitem{he2009efficient}
Z.~He, A.~Cichocki, and S.~Xie, ``{Efficient method for Tucker3 model
  selection},'' \emph{Electronics Letters}, vol.~45, no.~15, pp. 805--806,
  2009.

\bibitem{he2010detecting}
Z.~He, A.~Cichocki, S.~Xie, and K.~Choi, ``{Detecting the Number of Clusters in
  n-Way Probabilistic Clustering},'' \emph{IEEE Trans. on Pattern Analysis and
  Machine Intelligence}, vol.~32, no.~11, pp. 2006--2021, 2010.

\bibitem{huang2016bayesian}
L.~Huang, Y.~Xiao, K.~Liu, H.~C. So, and J.~Zhang, ``{Bayesian Information
  Criterion for Source Enumeration in Large-Scale Adaptive Antenna Array},''
  \emph{IEEE Trans. on Vehicular Technology}, vol.~65, no.~5, pp. 3018--3032,
  2016.

\bibitem{beheshti2018number}
S.~Beheshti and S.~Sedghizadeh, ``{Number of Source Signal Estimation by the
  Mean Squared Eigenvalue Error},'' \emph{IEEE Trans. on Signal Processing},
  vol.~66, no.~21, pp. 5694--5704, 2018.

\bibitem{suzuki2000detection}
M.~Suzuki, H.~Sanada, and N.~Naga, ``{Detection of signal number based on
  statistics of maximum likelihood},'' in \emph{Proc. of ICASSP}, vol.~2, 2000,
  pp. II733--II736.

\bibitem{wu2002determination}
Y.~Wu, K.-W. Tam, and F.~Li, ``{Determination of number of sources with
  multiple arrays in correlated noise fields},'' \emph{IEEE Trans. on Signal
  Processing}, vol.~50, no.~6, pp. 1257--1260, 2002.

\bibitem{chung2004detection}
P.~J. Chung, J.~F. Bohme, A.~O. Hero, and C.~F. Mecklenbrauker, ``{Detection of
  the number of signals using a multiple hypothesis test},'' in
  \emph{Processing Workshop Proceedings, 2004 Sensor Array and Multichannel
  Signal}, 2004, pp. 221--224.

\bibitem{choqueuse2008blind}
V.~Choqueuse, K.~Yao, L.~Collin, and G.~Burel, ``{Blind detection of the number
  of communication signals under spatially correlated noise by ICA and K-S
  tests},'' in \emph{Proc. of ICASSP}, 2008, pp. 2397--2400.

\bibitem{tu2010study}
S.~Tu and L.~Xu, ``{A study of several model selection criteria for determining
  the number of signals},'' in \emph{Proc. of ICASSP}, 2010, pp. 1966--1969.

\bibitem{rezaie2015determination}
R.~Rezaie and X.~R. Li, ``{Determination, separation, and tracking of an
  unknown time varying number of maneuvering sources by Bayes joint
  decision-estimation},'' in \emph{18th International Conference on Information
  Fusion (Fusion)}, 2015, pp. 1848--1855.

\bibitem{bazzi2016detection}
A.~Bazzi, D.~T.~M. Slock, and L.~Meilhac, ``{Detection of the number of
  superimposed signals using modified MDL criterion: A random matrix
  approach},'' in \emph{Proc. of ICASSP}, 2016, pp. 4593--4597.

\bibitem{szekely2007measuring}
G.~J. Sz{\'e}kely, M.~L. Rizzo, and N.~K. Bakirov, ``{Measuring and testing
  dependence by correlation of distances},'' \emph{{The Annals of Statistics}},
  vol.~35, no.~6, pp. 2769--2794, 2007.

\bibitem{andrews1992special}
L.~C. Andrews, \emph{{Special Functions of Mathematics for Engineers}}.\hskip
  1em plus 0.5em minus 0.4em\relax {McGraw-Hill New York}, 1992.

\bibitem{benesty2009pearson}
J.~Benesty, J.~Chen, Y.~Huang, and I.~Cohen, ``{Pearson correlation
  coefficient},'' in \emph{Noise reduction in speech processing}.\hskip 1em
  plus 0.5em minus 0.4em\relax Springer, 2009, pp. 1--4.

\bibitem{szekely2014partial}
G.~J. Sz{\'e}kely and M.~L. Rizzo, ``Partial distance correlation with methods
  for dissimilarities,'' \emph{{The Annals of Statistics}}, vol.~42, no.~6, pp.
  2382--2412, 2014.

\bibitem{huo2016fast}
X.~Huo and G.~J. Sz{\'e}kely, ``{Fast computing for distance covariance},''
  \emph{Technometrics}, vol.~58, no.~4, pp. 435--447, 2016.

\bibitem{cardoso1990eigen}
J.-F. Cardoso, ``{Eigen-structure of the fourth-order cumulant tensor with
  application to the blind source separation problem},'' in \emph{Proc. of
  ICASSP}, 1990, pp. 2655--2658.

\bibitem{hyvarinen1999fast}
A.~Hyvarinen, ``{Fast and robust fixed-point algorithms for independent
  component analysis},'' \emph{IEEE Trans. on Neural Networks}, vol.~10, no.~3,
  pp. 626--634, 1999.

\bibitem{franklin2012matrix}
J.~N. Franklin, \emph{{Matrix theory}}.\hskip 1em plus 0.5em minus 0.4em\relax
  Courier Corporation, 2012.

\bibitem{yeredor2010blind}
A.~Yeredor, ``{Blind Separation of Gaussian Sources With General Covariance
  Structures: Bounds and Optimal Estimation},'' \emph{IEEE Trans. on Signal
  Processing}, vol.~58, no.~10, pp. 5057--5068, 2010.

\bibitem{hitczenko1994rademacher}
P.~Hitczenko and S.~Kwapie{\'n}, ``{On the Rademacher series},'' in
  \emph{{Probability in Banach Spaces, 9}}.\hskip 1em plus 0.5em minus
  0.4em\relax Springer, 1994, pp. 31--36.

\end{thebibliography}
\bibliographystyle{unsrt}

\end{document}